\documentclass[10pt]{article}
\usepackage{latexsym,graphicx}
\usepackage{latexsym,graphicx}
\newcommand{\be}{\begin{equation}}
\newcommand{\ee}{\end{equation}}
\def\n{\noindent}
\catcode `\@=11 \catcode `\@=12
\begin{document}
\begin{center}
\large{\bf {Genesis of Dark Energy: Dark Energy as Consequence of
Release and Two-stage Tracking of Cosmological Nuclear Energy}}\\
\vspace{10mm}
\normalsize{R. C. Gupta $^1$ and Anirudh Pradhan $^2$} \\
\vspace{5mm} \normalsize{$^1$ Institute of Technology (GLAITM),
Mathura-281 406, India} \\
\normalsize{E-mail : rcg\_iet@hotmail.com, rcgupta@glaitm.org}\\
\vspace{5mm} \normalsize{$^2$ Department of Mathematics, Hindu
P. G. College, Zamania-232 331, Ghazipur, U. P., India} \\
\normalsize{E-mail : pradhan@iucaa.ernet.in}\\
\end{center}
\vspace{10mm}
\begin{abstract}
Recent observations on Type-Ia supernovae and low density
($\Omega_{m} = 0.3$) measurement of matter including dark matter
suggest that the present-day universe consists mainly of
repulsive-gravity type `exotic matter' with negative-pressure often
said `dark energy' ($\Omega_{x} = 0.7$). But the nature of dark
energy is mysterious and its puzzling questions, such as why, how,
where and when about the dark energy, are intriguing. In the present
paper the authors attempt to answer these questions while making an
effort to reveal the genesis of dark energy and suggest that `the
cosmological nuclear binding energy liberated during primordial
nucleo-synthesis remains trapped for a long time and then is
released free which manifests itself as dark energy in the
universe'. It is also explained why for dark energy the parameter $w
= - \frac{2}{3}$. Noting that $ w = 1$ for stiff matter and $w =
\frac{1}{3}$ for radiation; $w = - \frac{2}{3}$ is for dark energy
because $``-1"$ is due to `deficiency of stiff-nuclear-matter' and
that this binding energy is ultimately released as `radiation'
contributing $``+ \frac{1}{3}"$, making $w = -1 + \frac{1}{3} = -
\frac{2}{3}$. When dark energy is released free at $Z = 80$, $w =
-\frac{2}{3}$. But as on present day at $Z = 0$ when radiation
strength has diminished to $\delta \to 0$, $w = -1 +
\delta\frac{1}{3} = - 1$. This, thus almost solves the dark-energy
mystery of negative pressure and repulsive-gravity. The proposed
theory makes several estimates /predictions which agree reasonably
well with the astrophysical constraints and observations. Though
there are many candidate-theories, the proposed model of this paper
presents an entirely new approach (cosmological nuclear energy) as a
possible candidate for dark energy.
\end{abstract}
\smallskip
\n Key words : Cosmology, Dark energy, Nucleo-synthesis.\\\
\n PACS: 98.80.Ft, 98.80.Cq, 95.36.+x, 04.20.-q\\
\section{Introduction}
Recently observed astronomical phenomena have revolutionized the
understanding of cosmology. Consequences of combined analysis of Ia
Supernovae (SNe Ia) observations \cite{ref1,ref2}, galaxy cluster
measurements \cite{ref3} and cosmic microwave background (CMB) data
\cite{ref4} have shown that dark energy causing cosmic acceleration
dominates the present universe. This acceleration is observed at a
very small red-shift showing that it is a recent phenomenon in the
late universe. Observations of $16$ type Ia Supernovae made by
Hubble Space Telescope (HST) \cite{ref5} modified earlier
astronomical results and provided conclusive evidence for
deceleration prior to cosmic acceleration caused by dark energy in
the recent past. The acceleration is realized with negative pressure
and positive energy density that violates strong energy condition
(SEC) \cite{ref6}. This violation of SEC gives reverse gravitational
effect. Due to this effect, universe gets a jerk and transition from
deceleration to acceleration takes place. Phantom energy has
appeared as a potential dark energy candidate in this arena, which
violates both weak and strong energy conditions \cite{ref7,ref8}.
These data suggest $73\%$ content of the universe in the form of
dark energy (DE), $23\%$ in the form of non-baryonic dark matter
(DM) and the rest $4\%$ in the form of baryonic matter as well as
radiation. \\
\par
It is an irony of nature and is a puzzling phenomenon that most
abundant form of matter-energy in the universe is most mysterious.
The simplest dark energy candidate is the cosmological constant
$\Lambda$, but it needs to be extremely fine-tuned to satisfy the
current value of the dark energy. Alternatively, to explain decay of
the density, many dynamic models have been suggested, where
$\Lambda$ varies slowly with cosmic time (t) \cite{ref9,ref10}. In
addition to models with dynamic $\Lambda(t)$, many hydrodynamic
models with or without dissipative pressure have been proposed in
which barotropic fluid is the source of dark energy \cite{ref10}.
Chaplygin gas as well as generalized Chaplygin gas have also been
considered as possible dark energy sources due to negative pressure
\cite{ref8,ref11,ref12,ref13}. Other than these approaches, some
authors have considered modified gravitational action by adding a
function $f(R)$ (R being the Ricci scalar curvature) to
Einstein-Hilbert lagrangian, where $f(R)$ provides a gravitational
alternative for dark energy causing late-time acceleration of the
universe \cite{ref14,ref15,ref16,ref17}. A review on modified
gravity as an alternative to dark energy is available in
\cite{ref18}. A more comprehensive review is provided in
\cite{ref19}. All these models are phenomenological in the sense
that an idea of dark energy is introduced {\it a priori} either in
term of gravitational field or non-gravitational field. In spite
of these attempts, still cosmic acceleration is a challenge. \\
\par
Dark energy and the accelerated expansion of the universe have been
the direct prediction of the distant supernovae Ia observations
which are also supported, indirectly, by the observations of the CMB
anistropies, gravitational lensing and the studies of galaxy
clusters. It is generally believed that the distant SNe Ia
observations predict an accelerating expansion of the universe
powered by some hypothetical source with negative pressure generally
termed as `dark energy'. Supernovae at relatively high red shift are
found fainter than that predicted for an earlier-thought
slowing-expansion and indicate that expansion of universe is
actually speeding up \cite{ref1,ref2,ref20,ref21}. Recent studies
\cite{ref21,ref22} establish firmly that universe is now undergoing
an acceleration; with repulsive gravity of some strange energy-form
i.e., dark-energy at work. Dark energy, a `mysterious substance'
pressure of which is `negative' and accounts for nearly $70\%$ of
total matter-energy budget of the universe, but has no clear
explanation.  \\
\par
To understand the origin of dark energy and its nature is one of the
greatest problems of the 21st century. In the present paper, a
modest approach has been made to explain the genesis of dark energy,
suggesting that dark energy is a result of released-free
cosmological nuclear-energy in the universe; the origin of dark
energy lies within the nucleus. \\
\par
The out line of the paper is as follows. In section $2$, cosmo-
dynamics of FRW model is reviewed for completeness and further use
in the paper. Section $3$ presents a fresh approach towards
understanding the genesis of dark energy. Wherein we have suggested
that the dark energy is the released binding energy during the
primordial nucleo-synthesis. This dark energy decays in two stages:
in constraint way as stage-$1$ and liberated way in stage-$2$. In
section $4$, values of $\Omega_{x}$ at different values of Z have
been estimated and are found to be well in accordance with the
astrophysical constraints. In addition, theoretical estimates of
deceleration parameter, with conventional and modified method, have
been made in section $5$. After making critical discussions in
section $6$ about the genesis of dark energy, it is noticed that the
present model is in good agreement with the astrophysical
constraints. Finally, conclusions are summarized in section $7$.

\section{Cosmo-Dynamics}
The standard Friedman-Robertson-Walker (FRW) cosmological model
prescribes a homogeneous and an isotropic distribution for its
matter in the description of the present state of the universe. The
FRW equations and the analysis for cosmo-dynamics are briefly
summarized as follows for clarity and completeness sake and for its
subsequent uses:
\begin{equation}
\label{eq1} \left(\frac{\dot{a}}{a}\right)^{2} + \frac{k}{a^{2}} =
\frac{8\pi G}{3} \rho,
\end{equation}
\begin{equation}
\label{eq2} \frac{2\ddot{a}}{a} + \left(\frac{\dot{a}}{a}\right)^{2}
+ \frac{k}{a^{2}} = - 8\pi G p,
\end{equation}
where scale factor or universe-size at a particular time is $a(t)$
or simply as $a$. The dot denotes differentiation with respect to
cosmic time $t$.  Here, $G$ is gravitational constant, $p$ is
pressure and $\rho$ is the total density i.e., $\rho = \rho_{r} +
\rho_{m} + \rho_{x}$, where subscripts $r$ is for radiation, $m$ is
for matter (including dark matter), and $x$ is for dark energy. The
curvature factor is $k$ and depending on $k$ being $1, 0, -1$ the
universe is closed, flat or open. \\

Eqs. (\ref{eq1}) and (\ref{eq2}) lead to
\begin{equation}
\label{eq3} \frac{2\ddot{a}}{a} = - \frac{8\pi G}{3}(\rho + 3p),
\end{equation}
Differentiating Eq. (\ref{eq1}) and using (\ref{eq3}) in it, we get
\begin{equation}
\label{eq4} \dot{\rho} = -3(\rho + p)\frac{\dot{a}}{a}.
\end{equation}

Eq. (\ref{eq1}) can be written as
\begin{equation}
\label{eq5} H^{2} + \frac{k}{a^{2}} = \frac{8\pi G}{3} \rho,
\end{equation}
where $H$, being Hubble constant, is defined by
\begin{equation}
\label{eq6} H = \frac{\dot{a}}{a}.
\end{equation}
Critical density $\rho_{c}$ is defined as that for which the
universe is flat ($k = 0$), thus Eq. (\ref{eq5}) leads to
\begin{equation}
\label{eq7} \rho_{c} = \frac{3H^{2}}{8\pi G}.
\end{equation}
The deceleration parameter $q$ is defined as
\begin{equation}
\label{eq8} q = - \frac{a \ddot{a}}{\dot{a}^{2}} = -
\frac{\ddot{a}}{aH^{2}}.
\end{equation}
Using the equation of state parameter $w$ given by
\begin{equation}
\label{eq9} p = w\rho, ~ ~ \mbox{where} ~ ~  -1 \leq w \leq 1,
\end{equation}
in (\ref{eq4}), we obtain
\begin{equation}
\label{eq10} \frac{\dot{\rho}}{\rho} = - 3(1 + w)\frac{\dot{a}}{a},
\end{equation}
which leads to
\begin{equation}
\label{eq11} \rho \propto \frac{1}{a^{3(1 + w)}}.
\end{equation}
Eqs. (\ref{eq1}) and (\ref{eq11}), for flat universe, lead to
\begin{equation}
\label{eq12} a  \propto  t^{\frac{2}{3(1 + w)}}.
\end{equation}
Usually, a parameter called cosmological red-shift ($Z$) is often
referred to as a measure of the era (time t) of the universe. For
example; $Z_{0}$ corresponds to the present time $t_{0}$ with
present universe size (scale) $a_{0}$, and $Z$ corresponds to some
time t in the past with a smaller universe size a. In fact the
cosmological red-shift for spectral-line (wavelength $\lambda$) is
defined as $Z = \frac{d\lambda}{\lambda} = \frac{(a_{0} - a)}{a} =
\frac{a_{0}}{a} - 1$. Thus, $Z$ also gives a measure of universe
size at a particular time, higher $Z$ means smaller universe size in
the past and is expressed as
\begin{equation}
\label{eq13} \frac{a_{0}}{a} = 1 + Z.
\end{equation}
{\bf{Particular Cases:}} \\

(i) For $w = +1$, i.e., for stiff matter
\begin{equation}
\label{eq14} \rho \propto  \frac{1}{a^{6}} ~ ~ ~ \mbox{and} ~ ~ ~ a
\propto  t^{\frac{1}{3}}.
\end{equation}

(ii) For $w = + \frac{1}{3}$, i.e., for radiation dominated universe
\begin{equation}
\label{eq15} \rho \propto  \frac{1}{a^{4}} ~ ~ ~ \mbox{and} ~ ~ ~ a
\propto  t^{\frac{1}{2}}.
\end{equation}

(iii) For $w = 0$, i.e., for matter dominated (dust) universe
\begin{equation}
\label{eq16} \rho \propto  \frac{1}{a^{3}} ~ ~ ~ \mbox{and} ~ ~ ~ a
\propto  t^{\frac{2}{3}}.
\end{equation}

(iv) For $w = - \frac{2}{3}$, i.e., for dark energy (quintessence)
\begin{equation}
\label{eq17} \rho \propto  \frac{1}{a} ~ ~ ~ \mbox{and} ~ ~ ~ a
\propto t^{2}.
\end{equation}

(v) For $w = -1$, i.e., for vacuum energy.
\begin{equation}
\label{eq18} \rho = \mbox{constant} ~ ~ ~ \mbox{(like cosmolgical
constant)}.
\end{equation}

These variations `$\rho$ versus a' and `a versus t' of the
particular cases are easily obtained from Eqs. (\ref{eq11}) and
(\ref{eq12}) and are also shown in Figures $1$ and $2$.
\begin{figure}[htbp]
\centering
\includegraphics[width=11cm,height=14cm,angle=-90]{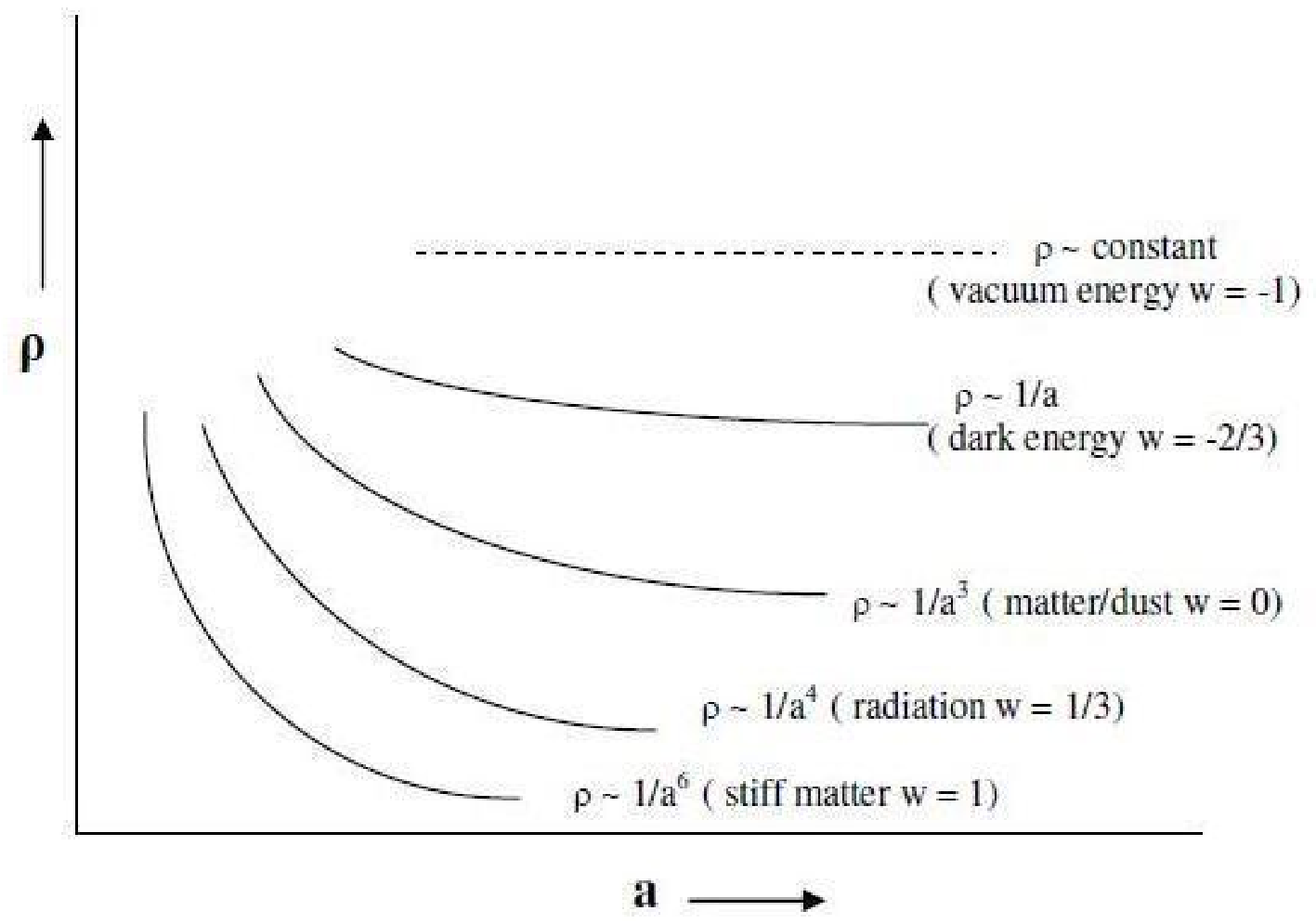} \\
\caption{Density $\rho$ decay with space size $a$}
\end{figure}
\begin{figure}[htbp]
\centering
\includegraphics[width=11cm,height=14cm,angle=-90]{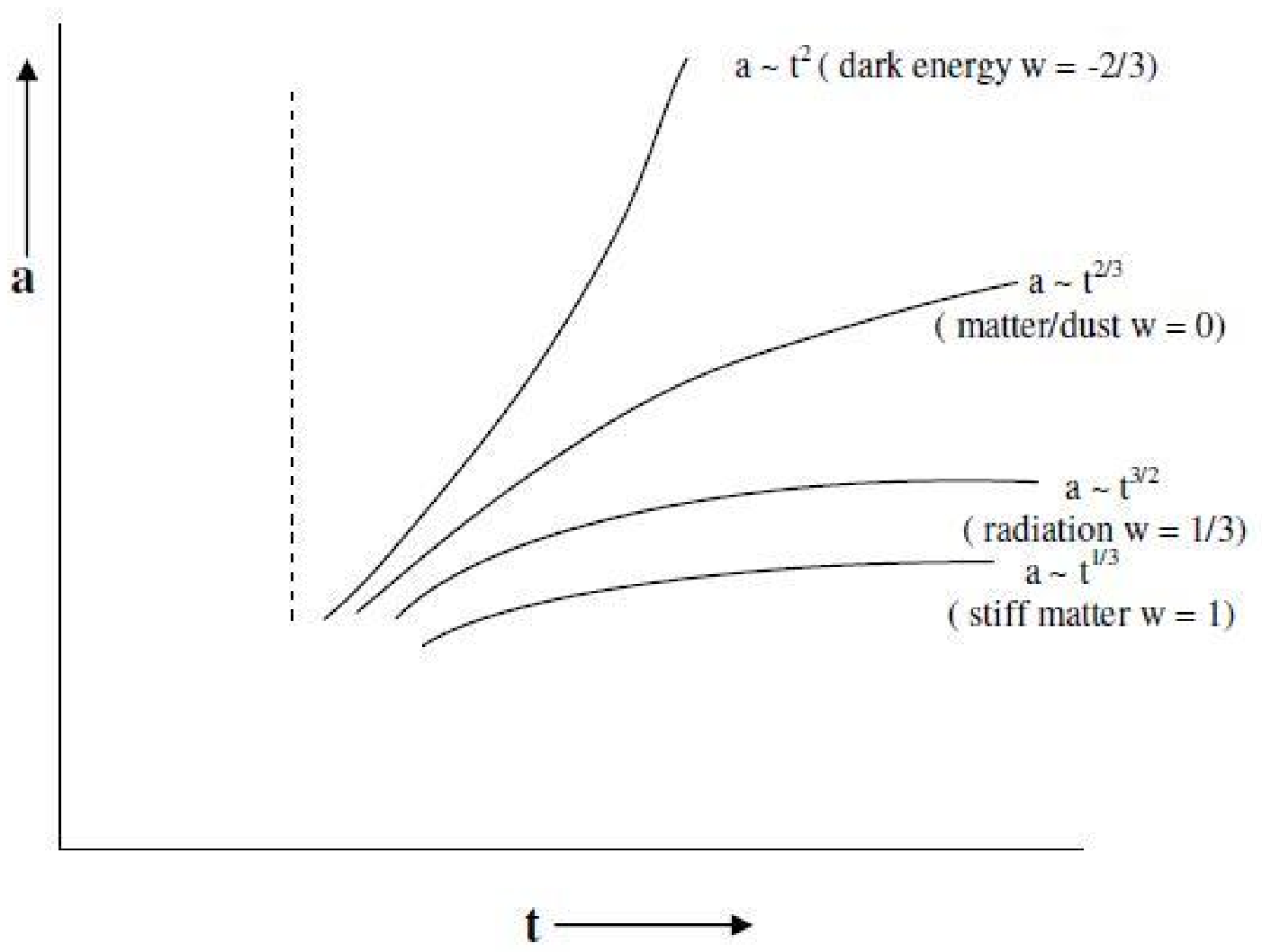} \\
\caption{Space size $a$ expansion with time $t$}
\end{figure}
\section{Nuclear Genesis of Dark Energy}

\subsection{Negative Pressure Due to Release of Cosmological Nuclear Energy}

It is suggested that dark energy is a consequence of nuclear energy.
The fusion nuclear binding energy begins to be liberated (but
trapped) during primordial nucleo-synthesis; and much later ($Z =
80$) after decoupling, the nuclear binding energy is released free
and appears effectively as dark energy. As shown (Figure $3$) in
subsequent section of the paper; this dark energy is produced
earlier but released free later, of course quantity-wise
($\Omega_{x} = 0.0003$) insignificant; remains low during galaxy
formation (say, at $Z = 3$, $\Omega_{x} = 0.1$), becomes $50\%$ at
transition ($Z = 0.5, \Omega_{x} = 0.5$) and emerges as dominant at
present time ($Z = 0, \Omega_{x} = 0.7$), satisfying all the
astrophysical
constants \cite{ref23}. \\

The equation of the state parameter $w$ for dark energy
(quintessence) is taken as $w = - \frac{2}{3}$. The reason for $w =
- \frac{2}{3}$ = $- 1 + \frac{1}{3}$ is explained as follows. Noting
that $w = + 1$ for {\it stiff} matter and $w = \frac{1}{3}$ for
radiation; `$-1$' is due to `deficiency of {\it stiff} (nuclear)
matter' which is manifested as mass-defect (binding-energy) and
`$\frac{1}{3}$' is because this released binding-energy though
initially-trapped but is finally released as `radiation'. Thus,
considering that the binding-energy due to mass-defect (deficiency)
in stiff-nucleus finally comes out as radiation; the equation of
state parameter $w = - 1 + \frac{1}{3} = - \frac{2}{3}$. When dark
energy is released free at $Z = 80$, $w = -\frac{2}{3}$. But as on
present day at $Z = 0$ when radiation strength has diminished to
$\delta \to 0$, $w = -1 + \delta\frac{1}{3} = - 1$, that is why dark
energy appears to be like the vacuum energy due to the cosmological
constant $\Lambda$ \cite{ref9,ref10}. The puzzle of negative
pressure (or repulsive gravity) is, thus, almost solved; the
negative pressure (or dark energy) is produced due to
released cosmological nuclear energy. \\
\newline
\begin{figure}[htbp]
\centering
\includegraphics[width=10cm,height=18cm,angle=0]{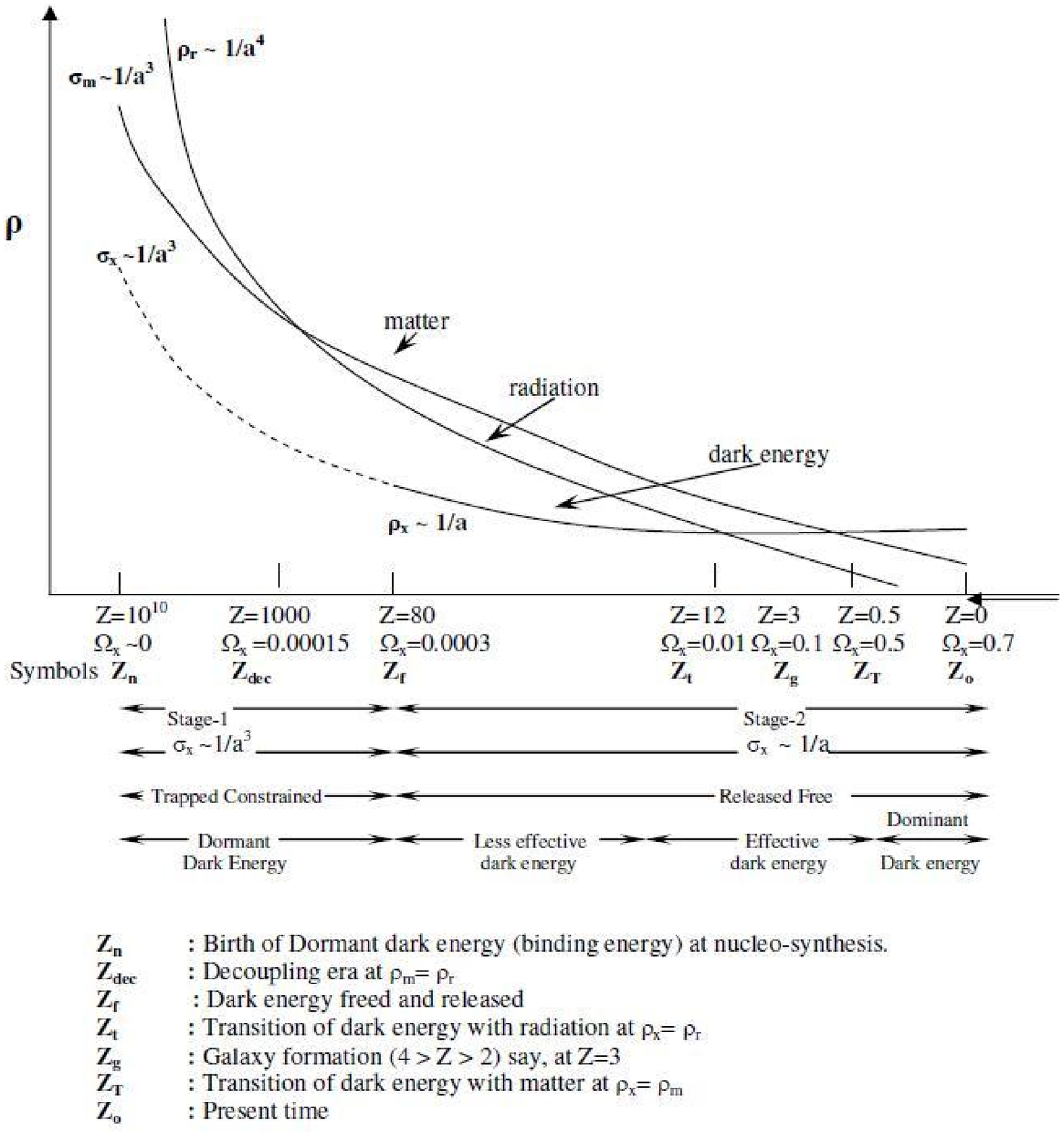} \\
\caption{Genesis of Dark Energy $\rho_{x}$ and its two stage
(Dominant and Free) variations/tracking along with variations of
$\rho_{r}$ and $\rho_{m}$}
\end{figure}
\subsection{Variation of Matter-Density and Dark-Energy-Density Much
After Decoupling Era Between $Z = 80$ to $Z = 0$}

Stiff-matter density decays very fast and vanishes very soon.
Radiation density also decays fast and vanishes soon much after
decoupling era. Although, as explained later in subsection $3.3$ and
in Figure $3$, the dark energy is born much earlier in the past
(nucleo-synthesis era) at about $Z = 10^{10}$ but remains trapped
dormant and decays in constrained way along with the matter and is
released free only after $Z = 80$ to be effective and dominant. For
flat universe ($\rho = \rho_{c}$), the expressions for the density
ratios can be written as follows.
\begin{equation}
\label{eq19} \rho = \rho_{c} = \rho_{r} + \rho_{m} + \rho_{x}
\end{equation}
and
\begin{equation}
\label{eq20} \Omega_{c} = \frac{\rho}{\rho_{c}}, ~ ~ \Omega_{r} =
\frac{\rho_{r}}{\rho_{c}}, ~ ~ \Omega_{m} =
\frac{\rho_{m}}{\rho_{c}}, ~ ~ \Omega_{x} =
\frac{\rho_{x}}{\rho_{c}}.
\end{equation}
Eqs. (\ref{eq19}) and (\ref{eq20}) lead to
\begin{equation}
\label{eq21} \Omega_{c} = 1 = \Omega_{m} + \Omega_{r} + \Omega_{x}.
\end{equation}
It is noticed    that between $Z = 80$ to $ Z = 0$, taking the
present values of densities $\rho_{m0} \approx  10^{-28}
gram/cm^{3}$, $ \rho_{x0} \approx   2\times 10^{-28} gram/cm^{3}$,
$\rho_{r0} \approx 10^{-31} gram/cm^{3}$ or $\Omega_{mo} \approx
0.3$ and $\Omega_{x0} \approx 0.7$ and negligibly small $\Omega_{r0}
\approx 0.0003$, the variations of radiation-density, matter-density
and dark-energy-density can be obtained, ratio-wise, from Eqs.
(\ref{eq15})$-$(\ref{eq17})) as follows.
\begin{equation}
\label{eq22} \rho_{r} = \frac{\rho_{r0}a_{0}^{4}}{a^{4}}=
0.0003\rho_{c}(1 + Z)^{4},
\end{equation}

\begin{equation}
\label{eq23} \rho_{m} = \frac{\rho_{m0}a_{0}^{3}}{a^{3}}=
0.3\rho_{c}(1 + Z)^{3},
\end{equation}
\begin{equation}
\label{eq24} \rho_{x} = \frac{\rho_{x0}a_{0}}{a}= 0.7\rho_{c}(1 +
Z),
\end{equation}
where the subscript $0$ indicates the value at present time ($Z =
0$).
\subsection{Biography of Dark Energy and Its Two-Stage Tracking}

Biography of mysterious dark-energy is even more mysterious.
Nevertheless, a brief biographical description of dark energy is
presented here for interesting clarity. \\

Dark energy in fact is the released nuclear binding energy of
cosmos. Dark energy is born during primordial nucleo-synthesis epoch
($Z \approx  10^{10}$) in the early universe out of `mother' matter
present there, while the nucleo-synthesis (of say, Helium) liberates
the binding energy as the `child' the child is trapped in or around
mother's lap and remains almost dormant for a long period and moves
parallel to mother's foot-steps. Approximate weight of this dormant
dark-energy (child) is roughly estimated as $1\%$ of binding energy,
of $25 \%$ of primordial helium-synthesis, from $13 \%$ ($0.13
\approx  \frac{0.04}{0.3}$) of baryonic-matter ($\Omega_{b} = 0.04$)
out of total mass of (mother) matter ($\Omega_{m} = 0.3$); this
comes out to be $\frac{\rho_{x}}{\rho_{m}} = 0.000325$ and since
this era is radiation dominated era $\rho_{r}$ being extremely high,
gives $\Omega_{x}$ very close to zero. This `child' i.e.,
dark-energy (nuclear-binding-energy liberated) though has come out
of mother's womb (nucleus) yet remains trapped/dormant for quite a
long time at least up to decoupling era $Z = 1000$ and even beyond.
When positive radiation ($\rho_{r}$) pressure is much more dominant
than the negative pressure of dark energy radiation ($\rho_{x}$),
the free-release of this dark energy is thus prohibited till then.
While the universe is expanding, the dark energy (the child) decays
parallel to matter (mother) foot-step at the same rate ($\rho
\propto  \frac{1}{a^{3}}$) from $Z = 10^{10}$ to $Z = 1000$ and even
beyond $Z< 100$ up to $Z = 80$ (Figure $3$). At $Z= 80$, though dark
energy $\Omega_{x}$ still being less than $\Omega_{r}$ (which
opposes liberation of the trapped dark energy) but $\Omega_{x}$ is
sufficient enough to fight with $\Omega_{r}$, the dark energy is
ultimately released. Thus at $Z = 80$, the dark energy (child) is
released free from mother's constrained-protection. Then onwards
($Z<80$) the matter density continues to decay at the same rate
($\frac{1}{a^{3}}$), but the dark energy (with $w = -\frac{2}{3} = -
1 + \frac{1}{3}$ at release time, as explained earlier in subsection
3.1) density decays at much slower rate ($\frac{1}{a}$) then
onwards. Between $Z = 80$ to $Z = 0$, the dark energy density curve
($\rho_{x} \propto \frac{1}{a}$) crosses radiation density curve
($\rho_{r} \propto \frac{1}{a^{4}}$) at $Z = 12$ and crosses matter
density curve ($\rho_{m} \propto \frac{1}{a^{3}}$) at $Z = 0.5$.
Briefly as summary (Figure $3$): `Dark energy is born as a result of
liberated binding energy during primordial nucleo-synthesis at $Z
\approx 10^{10}$ as a very small fraction
($\frac{\rho_{x}}{\rho_{m}} \approx  0.0003, \Omega_{x} \approx  0$)
of total matter-mass, initially (stage-1) it remains dormant and
decays fast ($\frac{1}{a^{3}}$) along with matter till $Z = 80$
($\frac{\rho_{x}}{\rho_{m}} \approx  0.0003, \Omega_{x} \approx
0.0003$) and finally it is released free (stage-2) and decays slowly
($\frac{1}{a}$) between $Z = 80$ to $Z = 0$ and becomes more
effective and even dominant today ($\frac{\rho_{x}}{\rho_{m}}
\approx  2, \Omega_{x} \approx  0.7$).' In stage-1: $\Omega_{x}$ is
almost equal to zero; whereas in stage-2: $\Omega_{x}$ increases as
$Z$ decreases; $\Omega_{x} = 0.0003$ at $Z = 80$; $\Omega_{x} =
0.01$ at $Z = 12$; $\Omega_{x} = 0.1$ at $Z = 3$; $\Omega_{x} = 0.5$
at $Z = 0.5$ and $\Omega_{x} = 0.7$ at $Z = 0$. The '{\it mantra}'
of dark energy hidden in nucleus comes out free and speaks aloud as
`$-1 + \frac{1}{3} = - \frac{2}{3}$', and orients itself towards
acceleration of the universe. Presently (as also mentioned in
section $3.1$), due to the fast decay of radiation strength, the
current value of $w$ could be close to `minus one'.

\section{Estimation of $Z$ and $\Omega_{x}$}

~ ~ (i) For present time; $Z = Z_{0} = 0$, $\Omega_{x} = 0.7$,
$\Omega_{m} = 0.3$ (Present-day known values). \\

(ii) For matter/dark-energy transition ($\rho_{m}=\rho_{x}$); Z can
be estimated from Eqs. (\ref{eq23}) and (\ref{eq24}) as $Z_{T} =
0.5$. \\

(iii) For galaxy formation (say at $Z = Z_{g} = 3$); $\Omega_{x}$
can be estimated from Eqs. (\ref{eq19}) $-$
(\ref{eq24}) as $\Omega_{xg} = 0.1$. \\

(iv) For radiation/dark-energy transition ($\rho_{r} = \rho_{x}$); Z
can be estimated from Eqs. (\ref{eq22}) and
(\ref{eq24}) as $Z = Z_{t} = 12$. \\

(v) At $Z = Z_{f} = 80$ when dark-energy is released 'free',
$\Omega_{x}$ can be estimated (in stage-2) from Eqs. (\ref{eq19})
$-$(\ref{eq24}) as $\Omega_{xf} = 0.0001$. The time $Z = 80$ is the
joining point for stage-1 and stage-2 (Figure $3$). In stage-1,
initial $\frac{\rho_{x}}{\rho_{m}} \approx  0.0003$ (see Section
$3.3$) remains constant as both $\rho_{x}$ and $\rho_{m}$ decay in
same way ($\frac{1}{a^{3}}$), and for stage-2
$\frac{\rho_{x}}{\rho_{m}} \approx \frac{0.7/0.3}{(1 + Z)^{2}}$ (see
Eqs. (\ref{eq23}) and (\ref{eq24})); equating these $0.0003 =
\frac{0.7/0.3}{(1 + Z)^{2}}$, gives $Z \approx  80$ as the meeting
point for the two stages. At the meeting point (${\rm {Z = 80}}$);
negative pressure of dark energy is much less than positive pressure
of matter and radiation energy, but appears to be strong enough so
as to enable the trapped dark energy to tunnel-through to come-out
as released-free.

(vi) At decoupling epoch ($\rho_{m} = \rho_{r}$) $Z = Z_{dec} =
1000, \Omega_{x}$ will be nearly half of the initial value of
$\frac{\rho_{x}}{\rho_{m}}$ in stage-1 (Figure $3$), i. e.,
$\Omega_{x}
= 0.00015$. \\

(vii) At about $ Z = Z_{n} \approx 10^{10}$ during nucleo-synthesis
era, birth of dark energy takes place. The fusion nuclear binding
energy ($ = 1\%$) begins to be liberated during primordial Helium
synthesis ($25\%$) from $13\%$ ($0.13 = \frac{0.04}{0.3}$) of
baryonic-matter ($\Omega_{b0} = 0.04$) out of total matter-mass
($\Omega_{m0} = 0.3$). Considering the liberated
nuclear-binding-energy as dark-energy, the dark-energy-density is
roughly estimated ($1\%$ of $25\%$ of $13\%$) at $0.000325$ of the
total matter-density. However, since radiation density at that high
value of Z (at nucleo-synthesis time $Z = Z_{n} \approx  10^{10}$)
would be much higher than that of matter, the estimated value for
$\Omega_{xn}$ would be
close-to-zero. \\

The liberated nuclear-energy as dark-energy, however, is trapped (as
mentioned in subsection 3.3 and in Figure $3$) in two stages. In
trapped stage-1 (between $Z = 10^{10}$ to $Z = 80$) the dark-energy
is trapped and dormant and varies ($\frac{1}{a^{3}}$) in a
constrained way along with its parent matter; whereas in free
stage-2 (between $Z = 80$ to $Z = 0$) the dark-energy is released
free and becomes effective/dominant and varies ($\frac{1}{a}$) at
slower rate during universe expansion. At $Z = 10^{10}$, $\Omega_{x}
\approx  0$;  at $Z = 80$, $\Omega_{x} = 0.0003$; at $Z = 12$,
$\Omega_{x} = 0.01$; at $Z = 3$, $\Omega_{x} = 0.1$; at $Z = 0.5$,
$\Omega_{x} = 0.5$ and at $Z = 0$, $\Omega_{x} = 0.7$. It may be
noted that these estimates are quite reasonable and satisfy
necessary astrophysical constraints \cite{ref23}; as shown in Figure
$4$, the density ratio $\Omega_{x}$ is initially too low and remains
low till galaxy formation.
\begin{figure}[htbp]
\centering
\includegraphics[width=11cm,height=17cm,angle=-90]{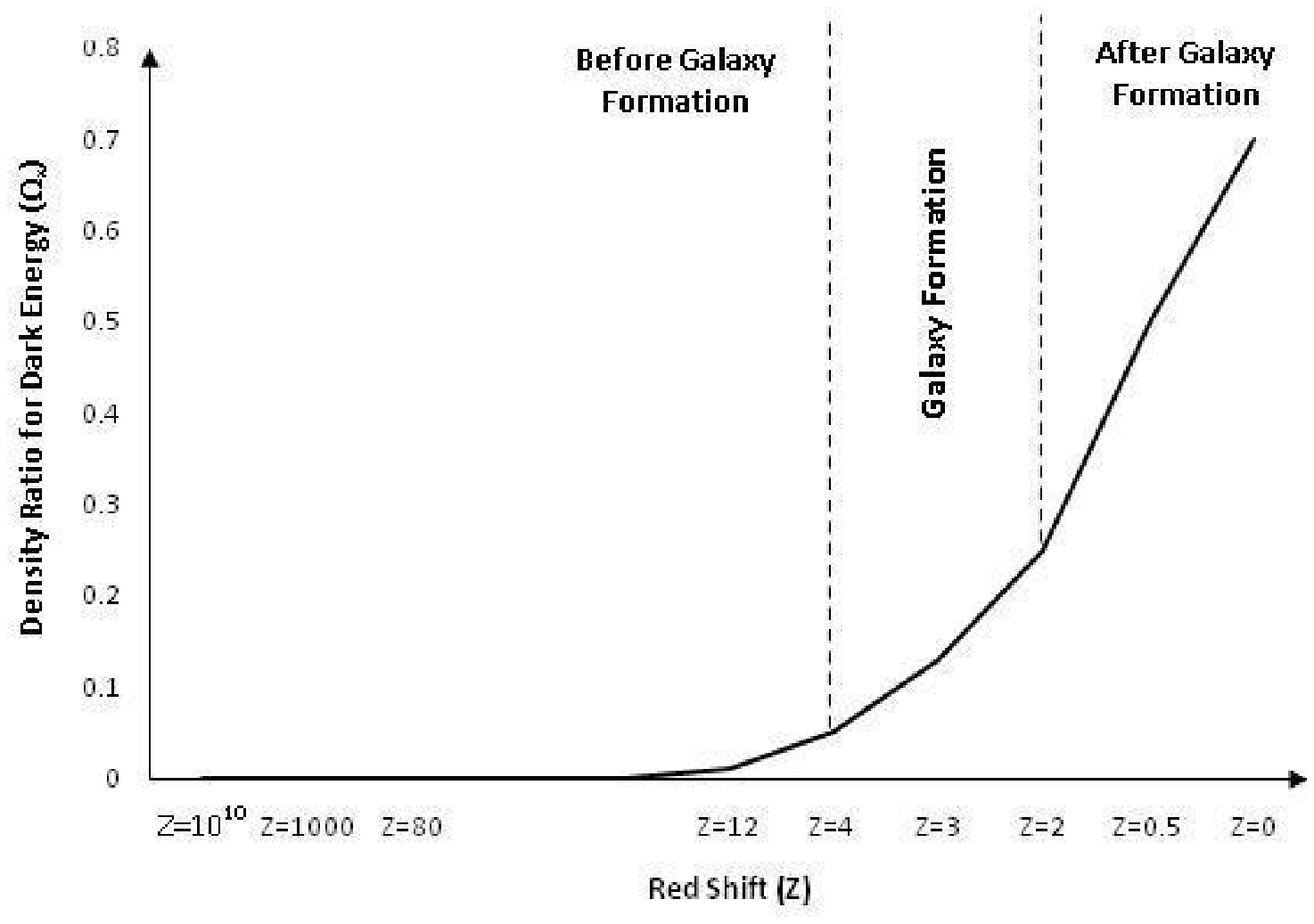} \\
\caption{Varation of density ratio for dark enery ($\Omega_{x}$)
verses Red shift (Z) showing that $\Omega_{x}$ is too low making
galaxy formation feasible}
\end{figure}

\section{Estimating Deceleration-Parameter $q$}

The space dimension for universe with (i) matter, (ii) dark-energy
and (iii) both these combined are written as follows:

(i) For matter ($w = 0$) universe
\begin{equation}
\label{eq25} a_{m} = a = K_{m} t^{\frac{2}{3}}, ~ ~ ~ \mbox{where} ~
~ ~ K_{m} = \frac{a_{0}}{t_{0}^{\frac{2}{3}}}.
\end{equation}

(ii) For dark-energy ($w = - \frac{2}{3}$) universe
\begin{equation}
\label{eq26} a_{x} = a = K_{x} t^{2}, ~ ~ ~ \mbox{where} ~ ~ ~ K_{x}
= \frac{a_{0}}{t_{0}^{2}}.
\end{equation}

(iii) For the universe combined with matter ($\Omega_{m}$) plus
dark-energy ($\Omega_{x}$)
\begin{equation}
\label{eq27} a = \Omega_{m} a_{m} + \Omega_{x} a_{x} = \Omega_{m}
K_{m} t^{\frac{2}{3}} + \Omega_{x} K_{x} t^{2}.
\end{equation}

The deceleration-parameter ($q = - \frac{a\ddot{a}}{\dot{a}^{2}}$)
can be estimated at present time $t_{0}$ ($\Omega_{m} = 0.3$,
$\Omega_{x} = 0.7$) using Eqs. (\ref{eq25}) $-$(\ref{eq27}) and is
found as $q_{0} = - 0.52$ (as also shown in subsection 5.2) which
indicates that the expansion of the universe is accelerating. The
current known value \cite{ref21} is, however, $q_{0} = - 0.67$. \\

In fact, there could be following two approaches to find the
deceleration-parameter which are described as follows:
\subsection{Conventional Method}
The `deceleration parameter' for `only matter' universe can easily
be evaluated as $q_{m} = \frac{1}{2}$ and that for `only
dark-energy' universe as $q_{x} = \frac{1}{2} + \frac{3}{2}w$. So a
simple conventional method usually followed assumes that the actual
`deceleration parameter $q$ for the matter plus dark-energy
universe' is `weighted sum of $q_{m}$ and $q_{x}$'. Thus,
\begin{equation}
\label{eq28} q_{0} = \frac{1}{2} \Omega_{m0} + \left(\frac{1}{2} +
\frac{3}{2}w\right)\Omega_{x0},
\end{equation}
which gives for $w = - \frac{2}{3}$, $\Omega_{m0} = 0.3$,
$\Omega_{x0} = 0.7$; $q = - 0.20$ a value much different from actual
known \cite{ref21} value of $q_{0} = -0.67$ necessitating
modifications in evaluation as follows.
\subsection{Modified Method}

The authors suggest a modified method to evaluate the
deceleration-parameter. This modified method assumes that actual
`space scale factor (a) for matter plus dark-energy universe' is
`weighted sum of $a_{m}$ and $a_{x}$'. Thus, the equation for total
actual space-size (a) can be written (similar to Eq. (\ref{eq27}))
as
\begin{equation}
\label{eq29} a = \Omega_{m} a_{m} + \Omega_{x} a_{x} = \Omega_{m}
K_{m} t^{\frac{2}{3}} + \Omega_{x} K_{x} t^{\frac{2}{3(1 + w)}}.
\end{equation}
The deceleration parameter ($q = - \frac{a\ddot{a}}{\dot{a}^{2}}$)
is evaluated, using Eqs. (\ref{eq25}) and (\ref{eq12}) in  Eq.
(\ref{eq29}) at present time $t_{0}$ as
\begin{equation}
\label{eq30} q_{0} = -\frac{[\frac{2}{9}\Omega_{m0} - \frac{2(1 + 3
w)}{9(1 + w)^{2}}\Omega_{x0}]}{[\frac{2}{3}\Omega_{m0} +
\frac{2}{3(1 + w)}\Omega_{x0}]^{2}},
\end{equation}
which gives for $w = - \frac{2}{3}$, $\Omega_{m0} = 0.3$,
$\Omega_{x0} = 0.7$; $q_{0} = -0.52$ (as found before using Eq.
(\ref{eq27})) which is reasonably near to the known \cite{ref21}
observed value of $q_{0} = - 0.67$. \\

Modified equation (\ref{eq30}) seems to be more correct than the
conventional method's expression given by Eq. (\ref{eq28}) since the
expansion (accelerating) of the universe due to dark-energy ($a \sim
t^{2}$) is much faster than the expansion (decelerating) of universe
due to matter ($a \sim t^{\frac{2}{3}}$). This fact is taken into
account in the modified-method (in Eqs. (\ref{eq27}), (\ref{eq29})
and (\ref{eq30}); the second term is more dominant). So, the
modified method predicts more nearer to value of $q_{0} = - 0.52$,
instead of $-0.20$ from conventional method, to the known observed
value of $-0.67$.
\section{Discussions}
The exact time of birth $Z = 10^{10}$ or $10^{9}$ cannot be
specified very precisely. It is suggested by the authors that
nuclear binding energy manifests itself as dark energy but the
beginning of nuclear energy and its completion is not exactly
defined/known. There is two-stage tracking (decay) of dark energy
(Figure $3$). Initially in stage-1 the decay is constrained and fast
during which the dark energy remains almost trapped/dormant for a
long time. Finally in stage-2 the dark energy is released free at $Z
= 80$ and decays slow and thus becomes more effective and dominant
in due course. The free released nuclear binding energy plays its
role as dark energy (with $w = - \frac{2}{3}$) creating
negative-pressure or repulsive-gravity. Due to the fast decay of the
radiation strength, the current value of $w$, however, could be $w =
-1$; indicating that as if vacuum energy is dark energy. Dark energy
begins with close-to-zero value of $\Omega_{x}$, remains low till
galaxy formation ($Z = 3, \Omega_{x} = 0.1$), becomes $50\%$ at
transition ($\Omega_{x} = 0.5, Z = 0.5$) and thereafter leads to
accelerated-universe becomes dominant at present time ($Z = 0,
\Omega_{x} = 0.7$). All these stand in accordance with the
astrophysical constraints \cite{ref23}. the dark-energy variations,
in stage-2 ($\rho_{x} \propto \frac{1}{a}$) is governed by
final/present condition (at $Z = 0$) and in stage-1 ($\rho_{x}
\propto \frac{1}{a^{3}}$) is governed by initial conditions (at $ Z
= 10^{10}$). The meeting point for the two stages (dotted and firm
curves in Figure $3$) is found at $Z = 80$. There could be some
error in it ($Z = 80$) due to approximations. Though the meeting
point is a point of interest in view of the fact that at this time
the dark-energy is finally released-free, however, any future change
in its estimate (other than $Z = 80$) will hardly have any effect on
other values (estimated in section 4) or on the theme
and philosophy of the model which will remain unchanged. \\

For estimating the deceleration parameter, the authors suggest a
more realistic modified-method (Eq. (\ref{eq30})) which gives an
estimate of $q_{0} = - 0.52$ against the observed known value of
$q_{0} = - 0.67$. The agreement is not bad. The discrepancy in the
value, however, may be due to following reason: the dark energy as a
result of liberation of nuclear binding energy not only takes place
during primordial nucleo-synthesis ($Z = 10^{10}$ or $10^{9}$) era
but also due to fusion reaction taking place within stars after or
around galaxy formation ($Z = 4$ to $0$). It may be noted that only
`primordial nucleo-synthesis', which indeed is more significant, has
been taken into consideration; but the `star nucleo-synthesis' will
also have some effect on acceleration of universe i.e., on the
deceleration parameter in the right direction. Also, this means that
nucleo-synthesis is more or less a continuous process. \\

Though there are many suspects (candidates) such as
\cite{ref24,ref25} cosmological constant, vacuum energy, scalar
field, brane world etc. as reported in the vast literature for the
dark energy, the proposed model in this paper at least presents a
new candidate (cosmological nuclear-energy) as a possible suspect
(candidate) for the dark energy.
\section{Conclusions}
The authors propose a novel model for genesis of dark energy,
indicating its origin in nucleus and suggesting that it is the
released-free nuclear binding energy of cosmos which manifests
itself as dark energy causing negative pressure and repulsive
gravity in the universe. It explains why for dark energy $w = -
\frac{2}{3}$ at release time and $w = -1$ at present time. Also, the
model describes the biography of the dark energy and tells that
$\Omega_{x}$ begins with a close-to zero value, remains low as
necessary till galaxy formation say at $Z = 3$, $\Omega_{x} = 0.1$;
transition at $Z = 0.5$, $\Omega_{x} = 0.5$ and becomes dominant at
present time $Z = 0$, $\Omega_{x} = 0.7$ satisfying very well all
the astrophysical constraints. The authors also propose a
modified-method to estimate deceleration parameter and find that
with the proposed model $q_{0} = -0.52$ in reasonable agreement with
known observed value. It can be metaphorically said that `dark
energy is a fossil of nuclear reaction that had taken place in the
early universe and that the fossil has recently been noticed when
the matter-cover over it has diminished in due course'. The secret
of acceleration of big-universe is hidden in the small-nucleus.
\section*{Acknowledgements}
The authors thank to the Institute of Engineering and Technology
(IET) and U. P. Technical University (UPTU), Lucknow, India for
providing facility where the part of this work was done. The authors
also thank Sushant Gupta for assistance in drawing the figures. The
financial support (Grant No.F.6-2(29)/2008(MRP/NRCB) of University
Grants Commission, New Delhi is gratefully acknowledged.

\end{document}